# Parabolic Vector Focus Wave Modes

Pavel Gotovski and Sergej Orlov

*Center for Physical Sciences and Technology, Industrial laboratory for photonic technologies,
Saulétekio ave 3, LT-10257 Vilnius, Lithuania
E-mail: sergejus.orlovas@ftmc.lt*

Weber-type parabolic beams have a transverse intensity profile, which is parabolically-shaped and can be flexibly controlled. On the other hand, this type of beams belongs to the family of the so-called nondiffracting beams and have properties, promising for applications where the shape of the beam is of an importance. Vector electromagnetic theory has to be introduced in order to fully describe optical beams inside a high numerical aperture system, where the angles of the spatial spectra are large. We introduce here parabolic vector focus wave modes (FWM), which are both resistant to diffraction and to material dispersion. We employ here a spectral approach and investigate durations of parabolic vector FWM's in the femtosecond region. Two cases of transverse electric and transverse magnetic modes are introduced and both standing and traveling types of waves are considered. We demonstrate how the angular dispersion affects the pulse shape and its properties. Parabolic vector FWM's are studied in a transparent dielectric media (sapphire), which is widely used as laser processed material.
DOI: 10.2961/jlmn.2019.01.0005

## 1. Introduction

Spatial and temporal localization of light energy is of great importance for various applications. The diffraction-free and dispersion-free propagation of pulsed beams (usually called as X waves) can be achieved both in linear [1–3] and nonlinear media [4, 5]. Such pulses are polychromatic superposition of nondiffracting Bessel beams, known under names like X-waves [6], Focus Wave modes [7] etc.

On the other hand, the so-called Weber-type parabolic beams have controllable parabolically-shaped transverse intensity profile distribution [8], which has promising applications in laser precise microfabrication. Also, Weber beams exhibit nondiffracting properties, what also greatly improves their practical applications. When a large spatial angles contribute to angular spectrum, vectorial electromagnetic theory must be introduced in order to fully describe laser beams. Monochromatic parabolic beams and their combinations are already represented in literature in great detail, however their pulsed beam counterparts still are underrepresented.

In this work we construct parabolic vector focal wave modes from scalar Weber-type parabolic wave solution [8, 9] using spectral approach [10, 11] and investigate durations up to tens of femtoseconds. We analyze two cases of transverse electric and transverse magnetic modes and consider waves, which are transversally standing or travelling. The impact of angular dispersion magnitude and type (positive or negative) on the resulting pulse shape is demonstrated. Moreover, the propagation of pulses not only in free space but also in a transparent dielectric media (sapphire) is analyzed.

## 2. Theory

We define parabolic cylinder coordinates by the transformation

$$x + iy = \frac{(\eta + i\xi)^2}{2}, \quad z = z. \qquad (1)$$

In these coordinates the three-dimensional Helmholtz equation separates into a longitudinal and transverse parts. Longitudinal part has solution with dependence $\exp(ik_z z)$, and a transverse part is a product $U(\eta,\xi,a) = R(\xi)\Phi(\eta)$, where scalar functions $R(\xi)$, $\Phi(\eta)$ depend on transverse coordinates $(\eta, \xi)$ and obey [8,9]

$$\frac{d^2\Phi(\eta)}{d\eta^2} + (k_t^2\eta^2 + 2k_t a)\Phi(\eta) = 0, \qquad (2)$$

$$\frac{d^2 R(\xi)}{d\xi^2} + (k_t^2\xi^2 - 2k_t a)R(\xi) = 0. \qquad (3)$$

Here $k_t$ is a transverse vector and $a$ is a parabolicity parameter. We change variables according to the rules $\sigma\xi -> v$ and $\sigma\eta -> u$, where $\sigma = \sqrt{2k_t}$, so Eqs. (2) and (3) are transformed into the canonical form of the parabolic cylinder differential equation

$$\frac{d^2 P}{dv^2} + (v^2/4 - a)P = 0 \qquad (4)$$

Function $P$ denotes here solutions of the parabolic differential equation and are found by standard methods (e.g., Frobenius), and their Taylor expansion at $v = 0$ is given by [8, 9]:

$$P(v,a) = \sum_{n=0}^{\infty} c_n \frac{v^n}{n!}, \quad c_{n+2} = ac_n - \frac{n(n-1)c_{n-2}}{4}. \qquad (5)$$

The transverse profiles $U(\eta,\xi,a) = R(\xi)\Phi(\eta)$ can be subdivided into even ($e$), odd ($o$) and traveling (t) non-diffracting parabolic beams with following transverse parts [7]:

$$U_e(\eta,\xi,a) = \frac{1}{\pi\sqrt{2}}|\Gamma_1|^2 P_e(\sigma\xi,a)P_e(\sigma\eta,-a), \quad (6)$$

$$U_o(\eta,\xi,a) = \frac{2}{\pi\sqrt{2}}|\Gamma_3|^2 P_o(\sigma\xi,a)P_o(\sigma\eta,-a), \quad (7)$$

$$U_t(\eta,\xi,a) = U_e(\eta,\xi,a) \pm iU_o(\eta,\xi,a), \quad (8)$$

where $\Gamma_1 = \Gamma(1/4+ia/2)$ and $\Gamma_3 = \Gamma(3/4+ia/2)$. We note that the odd and even type parabolic beams have not only positive but also negative $k_x$ and $k_y$ components in their spatial spectrum, therefore they do represent standing waves. On the other hand the travelling wave solution has either only positive $k_x$ or only positive $k_y$ (depending on the orientation of the coordinate system) In order to fully explain properties of nondiffracting pulses, full vectorial description must be used. Thus we vectorize scalar parabolic nondiffracting fields using

$$\mathbf{M}(\mathbf{r},\omega,a) = \nabla \times [\mathbf{e}_z U(\eta,\xi,a)\exp(ik_z z)], \quad (9)$$

$$k\mathbf{N}(\mathbf{r},\omega,a) = \nabla \times \mathbf{M}(\mathbf{r},\omega,a), \quad (10)$$

where $\mathbf{e}_z = (0,0,1)$ and $\omega$ is angular frequency. An azimuthally polarized parabolic Weber beam is represented here by a vector field $\mathbf{M}$ and the radially – by a vector field $\mathbf{N}$.

$$\mathbf{M}(\mathbf{r},\omega,a) = \vec{e}_x \left\{ \frac{\sigma e^{izk_z}\left(\xi U^{(0,1)}(\xi\sigma,\eta\sigma) + \eta U^{(1,0)}(\xi\sigma,\eta\sigma)\right)}{\eta^2 + \xi^2} \right\}$$
$$+ \vec{e}_y \left\{ \frac{\sigma e^{izk_z}\left(\eta U^{(0,1)}(\xi\sigma,\eta\sigma) - \xi U^{(1,0)}(\xi\sigma,\eta\sigma)\right)}{\eta^2 + \xi^2} \right\}, \quad (11)$$

$$\mathbf{N}(\mathbf{r},\omega,a) = \vec{e}_x \left\{ \frac{i\xi\sigma e^{izk_z} k_z U^{(1,0)}(\xi\sigma,\eta\sigma)}{k(\eta^2+\xi^2)} - \frac{i\eta\sigma e^{izk_z} k_z U^{(0,1)}(\xi\sigma,\eta\sigma)}{k(\eta^2+\xi^2)} \right\}$$
$$+ \vec{e}_y \left\{ \frac{i\xi\sigma e^{izk_z} k_z U^{(0,1)}(\xi\sigma,\eta\sigma)}{k(\eta^2+\xi^2)} + \frac{i\eta\sigma e^{izk_z} k_z U^{(1,0)}(\xi\sigma,\eta\sigma)}{k(\eta^2+\xi^2)} \right\}$$
$$+ \vec{e}_z \left\{ \frac{i\sigma e^{izk_z}\left(i\sigma U^{(0,2)}(\xi\sigma,\eta\sigma)+i\sigma U^{(2,0)}(\xi\sigma,\eta\sigma)\right)}{k(\eta^2+\xi^2)} \right\}, \quad (12)$$

where $U^{(n,k)}(\xi\sigma,\eta\sigma) = \dfrac{\partial^{n+k} U(\xi\sigma,\eta\sigma)}{\partial \xi^n \partial \eta^k}$.

The model of so called Focus wave modes [6] is general in the sense that it describes pulses without any transversal or longitudinal dispersion using any frequency spectrum of the source and strictly predetermined angular dispersion $\theta = f(\omega)$ of individual Bessel beams given inside linear medium by [3, 6, 12]

$$k_z(\omega) = \frac{\omega}{c}n(\omega)\cos\theta = \frac{\omega}{V} + \gamma. \quad (13)$$

In the ideal case FWMs propagate in the air or inside a dispersive medium and its group velocity remains constant over the whole spectral range $dk_z/d\omega = 1/V$ and the parameter $\gamma$ is an integration constant, which influences the phase velocity, see [6,12]. Thus, the longitudinal projection $k_z$ of the wave vector $\mathbf{k}$ must satisfy relation (13).

Nondiffracting vector parabolic pulsed beams are introduced here as a polychromatic superposition of vector Weber parabolic beams. Thus, vector parabolic focus wave modes are described as

$$\mathbf{E}(\mathbf{r},t) = \int_0^\infty S(\omega)\mathbf{E}_0(\mathbf{r};\omega)\exp(-i\omega t)d\omega, \quad (14)$$

where $\mathbf{E}_0(\mathbf{r},\omega)$ is a complex amplitude of vector monochromatic beam. For the cases of azimuthally or radially polarized parabolic vector beams the FWMs can be obtained replacing $\mathbf{E}_0(\mathbf{r},)$ with expression $\mathbf{M}$ and $\mathbf{N}$ from Eqs. (9,10).

### 3. Simulation and numerical results

The typical angular dispersion curves in sapphire for different values of $V/c$ for a single value of $\gamma$ are shown in Fig. 1. The frequency $\omega$ is normalized to frequency $\omega_0$ corresponding to wavelength $\lambda = 1.028\,\mu m$, which is the carrier wavelength of femtosecond laser system in the lab ("Pharos" by Light Conversion). We fix the parameter $\gamma$ at $\gamma = -5.067\times10^5$ m$^{-1}$ as it ensures a rather rich choice of different angular dispersion curves for different values of $V/c$.

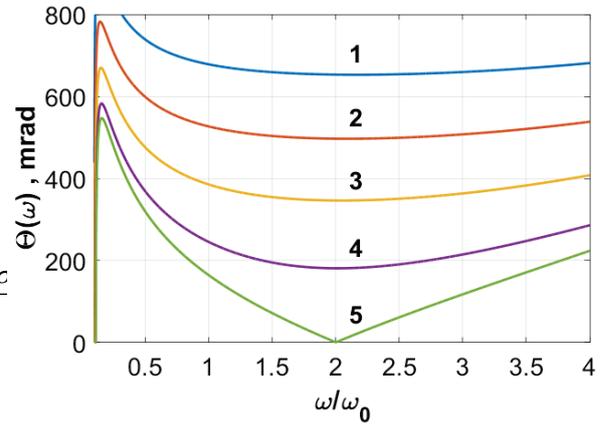

**Fig. 1** Dispersion curves of diffraction free parabolic pulses in Sapphire. Parameter $\gamma=-5.067\times 10^5$ m$^{-1}$, V/c=0.69 (1), V/c=0.625 (2), V/c=0.585 (3), V/c=0.56 (4), V/c=0.5510874 (5), $\omega_0=1.83\times10^{15}$s$^{-1}$.

Before we proceed with discussion on our numerical results we note, that sapphire is a birefringent uniaxial material. The difference between refractive indices of ordinary and extraordinary beams in our region of interest is $\Delta n = 10^{-5}$. We choose the direction of the crystal axis that coincides with the axis of propagation $z$. In this case the azimuthal polarization is the ordinary beam and the radial polarization is the extraordinary beam.

For the sake of brevity we restrict ourselves with the case of a radially polarized FWMs, described by the Eq. (9, 12). We note, that for angles and wavelengths depicted in Fig. 1 the refractive index of the extraordinary beam is equal to the refractive index of the ordinary beam and the birefringence of the sapphire can be neglected. We also integrate in Eq. (12) over a rectangular spectral amplitude $S(\omega) = \Pi\left(2(\omega-\omega_c)/\Delta\omega\right)$ (here $\Pi$ is a boxcar function) with the center located at the carrier frequency $\omega_c$ and with spectral width $\Delta\omega$.

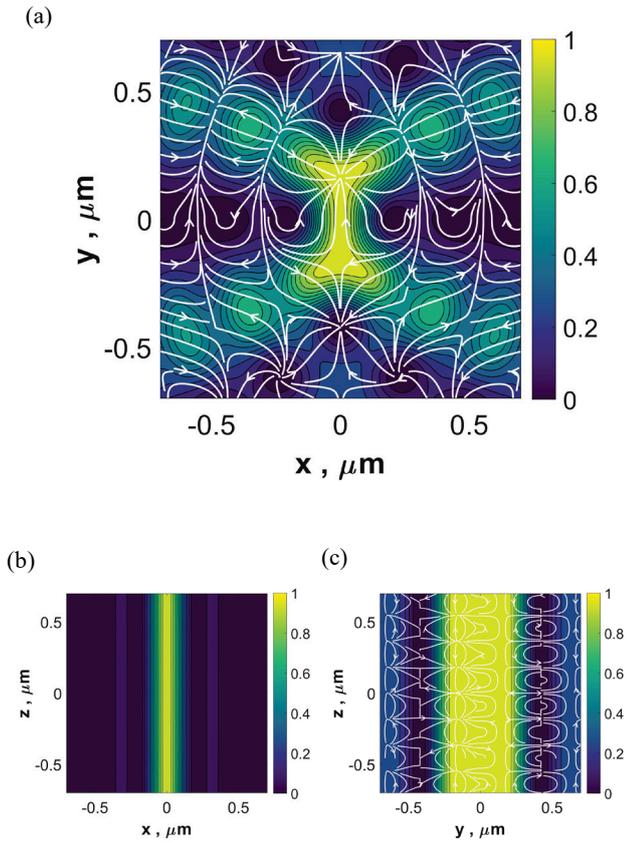

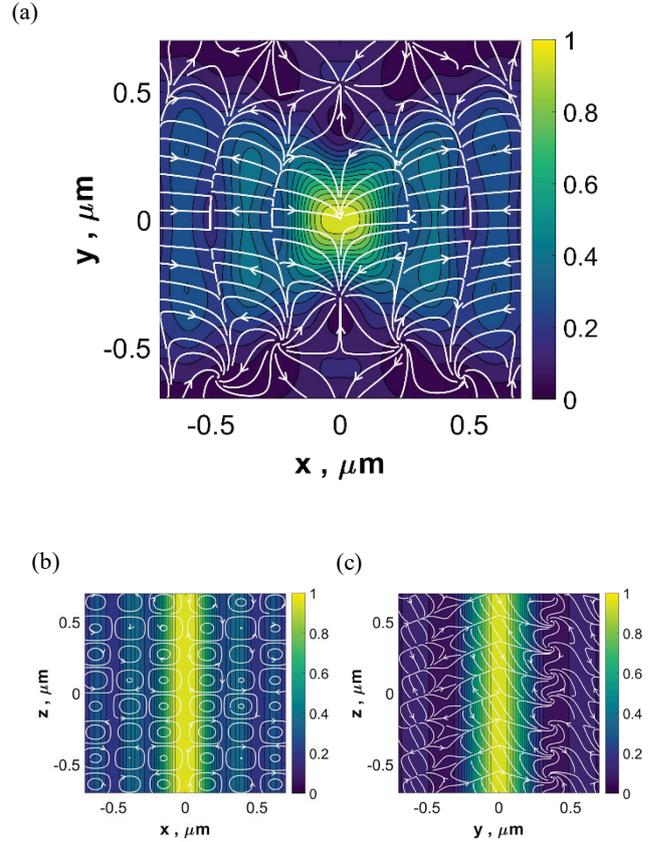

**Fig. 2** Intensity distribution of the odd radially polarized parabolic FWM inside sapphire for $a = 0$ in the tranverse $(x,y)$ plane (a), in the longtitudanal $(x,z)$ plane (b) and in the longtitudanal $(y,z)$ plane (c). Parameters are $\gamma=-5.067\times10^5$ m$^{-1}$, $V/c$=0.69, $\omega_0$=1.83×10$^{15}$s$^{-1}$, $\omega_c$=2$\omega_0$ and $\Delta\omega = 4\times10^{13}$ s$^{-1}$.

**Fig. 4** Intensity distribution of the travelling type radially polarized parabolic FWM inside sapphire for $a = 0$ in the tranverse $(x,y)$ plane (a), in the longtitudanal $(x,z)$ plane (b) and in the longtitudanal $(y,z)$ plane (c). Parameters are $\gamma=-5.067\times10^5$ m$^{-1}$, $V/c$=0.69, $\omega_0$=1.83×10$^{15}$s$^{-1}$, $\omega_c$=2$\omega_0$ and $\Delta\omega = 4\times10^{13}$ s$^{-1}$.

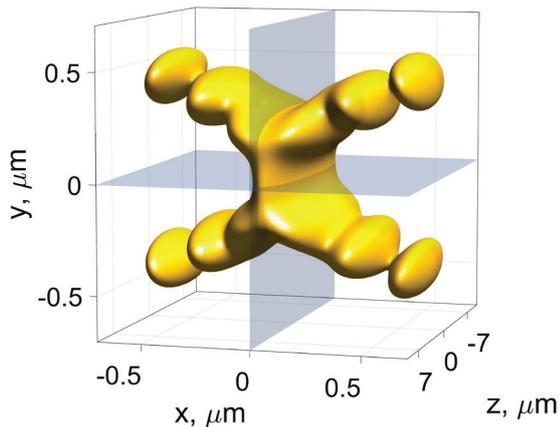

**Fig. 3** A three dimensional depiction of an intensity isosurface of the odd type radially polarized parabolic FWM inside sapphire for $a = 0$. Parameters are $\gamma=-5.067\times10^5$ m$^{-1}$, $V/c$=0.69, $\omega_0$=1.83×10$^{15}$ s$^{-1}$, $\omega_c$=2$\omega_0$ and $\Delta\omega = 4\times10^{13}$ s$^{-1}$. The isosurface corresponds approximately to 36% of the total intensity.

The expression of the electric field is given then by

$$\mathbf{E}(\mathbf{r},t) = \int_{\omega_c-\Delta\omega/2}^{\omega_c+\Delta\omega/2} S(\omega)\mathbf{N}(\mathbf{r},\omega,a)\exp(-i\omega t)d\omega. \quad (15)$$

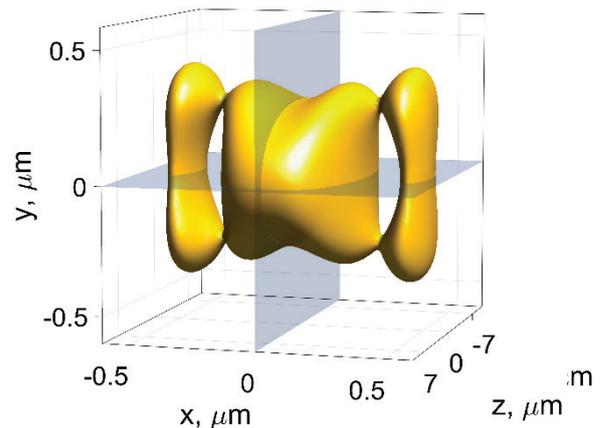

**Fig. 5** A three dimensional depiction of an intensity isosurface of the travelling type radially polarized parabolic FWM inside sapphire for $a = 0$. Parameters are $\gamma=-5.067\times10^5$ m$^{-1}$, $V/c$=0.69, $\omega_0$=1.83×10$^{15}$s$^{-1}$, $\omega_c$=2$\omega_0$ and $\Delta\omega = 4\times10^{13}$ s$^{-1}$. The isosurface corresponds approximately to 36% of the total intensity

The case of the azimuthally polarized parabolic FWMs will be presented elsewhere.

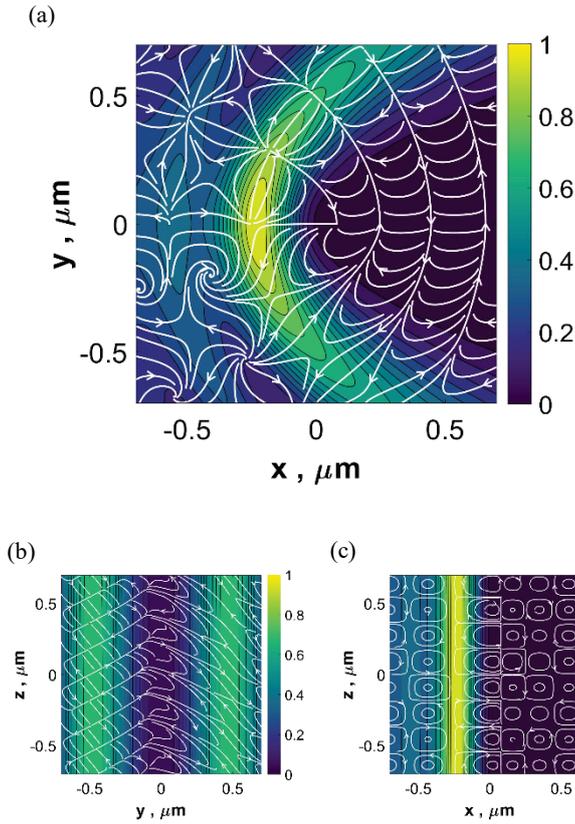

**Fig. 6** Intensity distribution of the travelling type radially polarized parabolic FWM inside sapphire for a = 2 in the tranverse (*x,y*) plane (a), in the longitudinal (*x,z*) plane (b) and in the longitudinal (*y,z*) plane (c). Parameters are $\gamma = -5.067\times 10^5$ m$^{-1}$, V/c=0.69, $\omega_0=1.83\times 10^{15}$s$^{-1}$, $\omega_c=2\omega_0$ and $\Delta\omega = 4\times 10^{13}$ s$^{-1}$.

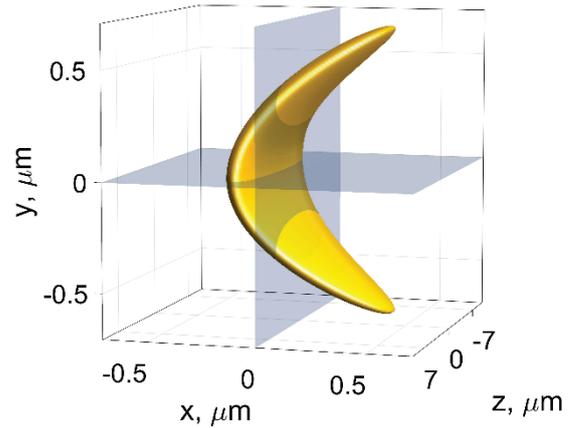

**Fig. 7** A three dimensional depiction of an intensity isosurface of the travelling type radially polarized parabolic FWM inside sapphire for for *a* = 2. Parameters are $\gamma=-5.067\times 10^5$ m$^{-1}$, V/c=0.69, $\omega_0=1.83\times 10^{15}$s$^{-1}$, $\omega_c=2\omega_0$ and $\Delta\omega = 4\times 10^{13}$ s$^{-1}$. The isosurface corresponds approximately to 36% of the total intensity.

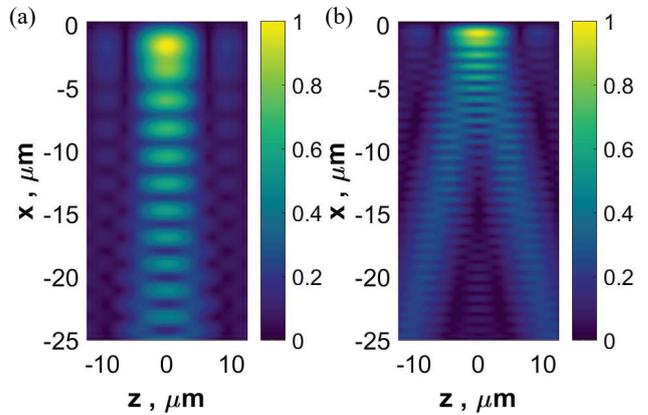

**Fig. 8** A comparison of the O type and the X type radially polarized parabolic FWM. Intensity distribution of the travelling type radially polarized parabolic FWM inside sapphire for a = 2 and $\Delta\omega = 4\times 10^{13}$ s$^{-1}$ in the longitudinal (*x,z*). For the O type $\omega_c=1.23\,\omega_0$ (a), for the X-type wave $\omega_c=3\omega_0$ (b).

We start with the case of zero parabolicity ( $a = 0$ ), we choose an odd type standing wave solution, see Eq. (7,12) and plot the electric field distributions in two planes (*x,y*) and (*x,z*), see Fig 2.

Additionally, we demonstrate here a three dimensional distribution of the intensity profile, see Fig 3. Obviously, as long as the parabolicity is zero, we observe a symmetrical shape.

As a next step, we investigate changes introduces by a switch from a standing type solution to a traveling type solution, see Eq. (8,12). Results for the same set of parameters are presented in Fig. 4. First, we see, that the travelling wave has a smoother transverse profile, which is free from the oscillations observed in the Fig. 2.

However, the shape of the central peak has changed also and it became more Gaussian-like. A three dimensional distribution of the intensity profile for this case is depicted in the Fig. 5.

Obviously, one can still find slight similarities between isosurfaces for the odd and travelling case, however the overall resemblance is low. Naturally, this case is also lacking any manifestation of the parabolic nature of the pulsed beams. However, as soon as we increase the parabolicity of radially polarized FWM, we start to notice a distinct parabolical shape in the transverse plane, see Fig. 6 (a-b).

First, we notice that the intensity has now a distinct parabolic shape and is shifted away from the center of the coordinates. The structure of the electric field in the vicinity of the center is slightly reminiscent to the structure of a radially polarized Bessel beam, see Ref. [13]. Another two cross-sections also demonstrate a distinct asymmetry in the vicinity of the center.

The complicated structure of the beam is surprisingly revealed in the three dimensional plot of the intensity isosurface, see Fig. 7.

Next, we discuss an influence of the dispersion curve behaviour on pulsed beam radius and its length. We choose now the angular dispersion curve with most drastic changes - the fifth angular dispersion curve from the Fig. 1 and keep the parabolicity equal to $a = 2$. Now, we select four different pulse durations (i.e. different spectral bandwidths $\Delta\omega$) and use Eq. (15) to analyze, how the carrier frequency $\omega_c$ influences both longitudinal and transverse dimensions of the pulsed beam.

It was noted, that for original X-waves and FWM's the shape of the spatio-temporal distribution undergoes changes inside the dispersive media, if the central carrier frequency is at the inflexion point with nearly zero angle of the plane waves cone, see Ref [12] and references therein. This situation is usually referred to as a transformation from the X letter shape, i.e. from the X-type wave, to the O letter shape (i.e. to the O-type wave). In the case of parabolic vector FWMs we do observe a similar behavior, see Fig. 8 for an example. Thus, neither parabolic character nor polarization properties do not influence this underlying behavior which seems to be more influenced by the dispersion of the FWM inside the dielectric material.

Next, we note, that in contrary to the conventional FWM's, which were in the great detail analyzed in Ref. [12], radially polarized parabolic FWMs have one distinct difference. Their spatial length $L_z$ (which we define as FWHM length of the intensity profile) does depend on the angular dispersion, see Fig. 9.

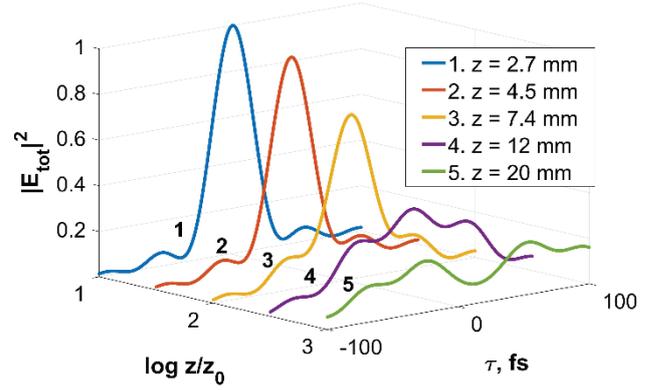

**Fig. 10**. Evolution of the pulse profile in the plane $y=0$ for the radially polarized parabolic FWM for $a = 2$ on the normalized distance $z/z_0$ from the air-glass interface ($z_0=1$ mm). Parameters are $\gamma=-5.067\times10^5$ m$^{-1}$, $V/c=0.5510874$, $\omega_0=1.83\times10^{15}$s$^{-1}$, $\Delta\omega = 8\times10^{13}$ s$^{-1}$, and $\omega_c= 2\ \omega_0$.

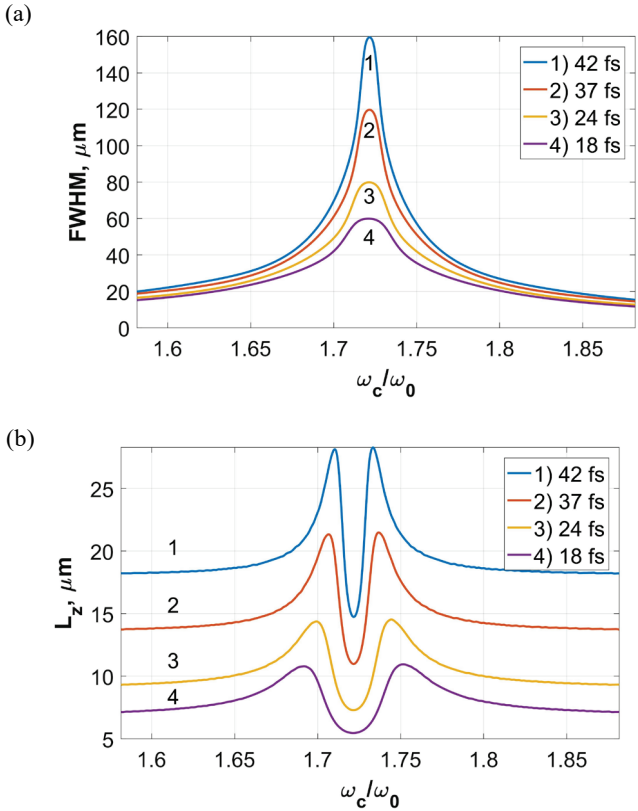

**Fig. 9**. Dependencies of the lengths $L_z$ (a) and FWHM transverse beam widths (b) of the radially polarized parabolic FWM inside sapphire for $a = 2$ on the normalized carrier frequency $\omega_c/\omega_0$. Parameters are $\gamma=-5.067\times10^5$ m$^{-1}$, $V/c=0.5510874$, $\omega_0=1.83\times10^{15}$s$^{-1}$.

This is caused by the fact, that the center of the pulsed beam, i.e. the on-axis point has no intensity, and the intensity is shifted away from the center in the $(x,y)$ plane to a point $(x',y')$ due to the polarization structure. Therefore, the length of the pulsed beam is evaluated not on the axis (i.e. in the point $x = 0$ and $y = 0$) but in the point with highest intensity $(x',y')$, so the expected pulse durations does not coincide with that of the sinc function.

Lastly, we would like to evaluate how the subtle angular dispersion relation, described by the Eq. (13) can be affected by the planar interface between the air and dielectric.

We assume, that the parabolic radially polarized FWM is created with the angular dispersion, which ensures dispersionless propagation in the air. Now, we use the formalism of the Fresnel coefficients from [10] and evaluate electric fields inside the sapphire. It's quite obvious, that the resulting field will experience distortions due to material dispersion. However, let us analyze how fast those effects occur for an radially polarized parabolic FWM inside the bulk of the dielectric medium. Results are depicted in the Fig. 10. Note, for the sake of brevity we don't plot here the dispersion curve.

Surprisingly enough, the femtosecond parabolic FWM pulse with initial duration of 35 fs experiences very little dispersion until the pulsed beam travels more than 7 mm away from the planar interface. Of course, afterwards it experience a quite rapid decay into a train of two pulses. Thus, for small angles and regions not far away from the interface, there is no need to introduce any additional compensation into the angular dispersion.

Nevertheless, a successful implementation of an additional dispersion for a FWM pulse in the air is straight – forward. At the boundary between the air and glass, the transverse wave vector projection remains the same for reflected and transmitted pulses, while longitudinal components change $k_{z1} \rightarrow k_{z2}$. We have implemented in the code this change, so it will end up with the proper angular dispersion inside the bulk of the transparent material (sapphire) and the parabolic FWM will preserve its duration while propagating inside the glass. Further results on this topic will be presented elsewhere.

## 4. Conclusions and discussions

Nonhomogeneous polarizations like azimuthal and radial are becoming important in areas of laser microprocessing, as the number of reports on their efficiency for material processing applications is increasing [14, 15]. Given their unique transverse parabolic shape and nondiffracting properties combined with conical nature vector parabolic shovel beams (focus wave modes) should be interesting for laser microprocessing applications, when round cuts and controllable corners are important.

We have studied here the propagation of radially polarized parabolic FWMs inside a transparent material (sapphire). We have revealed both the influence of the parabolicity parameter and of the carrier frequency on the shape of the parabolic azimuthally polarized focus wave mode. Lastly we have covered problems arising when a parabolic FWM enters the sapphire from the air through a planar interface. A dispersive pulse spreading is observed due to material dispersion and due to the change in the angular dispersion introduced by the interface.

**Acknowledgments and Appendixes**

This research is/was funded by the European Social Fund according to the activity 'Improvement of researchers' qualification by implementing world-class R&D projects' of Measure No. 09.3.3-LMT-K-712.